\documentstyle[preprint,aps]{revtex}

\begin{document}
\draft
\title{\Large Vacuum instability in  external fields}
\author{S.P. Gavrilov\thanks{On leave from Tomsk Pedagogical
University, 634041 Tomsk, Russia; present  
e-mail: 
gavrilov@snfma1.if.usp.br
} and D.M. Gitman\thanks{e-mail: 
gitman@snfma1.if.usp.br}} 
\address{Instituto de F\'{\i}sica, Universidade de S\~ao Paulo\\
P.O. Box 66318, 05389-970 S\~ao Paulo, SP, Brasil}
\date{\today}
\maketitle

\begin{abstract}
We study particles creation from the vacuum by 
external  electric fields, in particular,  by  fields, which are
acting for a  finite time,  in the
frame of QED in  arbitrary space-time dimensions.  In all the  cases  special sets of
exact solutions of Dirac equation (IN- and OUT- solutions) are
constructed. Using them, characteristics of the effect are
calculated. The time and dimensional analysis of the vacuum
instability is presented. It is shown  that
the distributions of particles created by quasiconstant electric
fields can be written in a form which has a thermal
character and seams to be universal, i.e. is valid for any theory with
quasiconstant external fields. Its application, for example,   to the
particles creation in external constant gravitational field reproduces
the Hawking temperature exactly.
\end{abstract}


\newpage

\section{Introduction}

The effect of particles creation  from  vacuum by an external field
(vacuum instability in an external field) ranks
among the most intriguing nonlinear phenomena in quantum theory. Its
consideration is theoretically important, since it requires one to go beyond
the scope of the perturbation theory, and its experimental observation
would verify the validity of the theory in the superstrong field domain. The
study of the effect began, in fact, first in $3+1$-dimensional QED in
connection with the so-called Klein \cite{b1} paradox, which revealed
the possibility of electron penetration through an arbitrary high
barrier formed by an external field. Then in 1951 Schwinger \cite{b2} found
the vacuum-to-vacuum transition probability in a constant electric field. It
became clear that the effect can  actually be
observed as soon as the external field strength approaches the
characteristic value (critical field)  $E_c=m^{2}c^{3}/|e|\hbar\simeq 1,3\cdot
10^{16}\; V/cm$.  Although  a real  possibility of creating such fields under 
laboratory conditions does not exist now, these fields can play a role
in astrophysics, where the
characteristic values of electromagnetic fields near  and 
 gravitational fields near black holes are enormous. One can also
mention that  Coulomb fields of superheavy nuclei can create 
electron-positron pairs. General considerations, concrete calculations
and detailed bibliography regarding the vacuum instability in QED can
be found in \cite{b9c,b3,b5,b6}.   Particles creation by
external gravitational fields \cite{b5,b4,b21} and non-Abelian
gauge fields \cite{b17}  can also
be considered in analogy with  electrodynamics. There are  also various problems in
modern quantum theory which are  closely related to the vacuum instability
in question, for example,  phase transitions in  field
theories,  the problem  of boundary conditions or topology
influence on the vacuum, the problem of consistent
vacuum construction in QCD, string theories, multiple particles
creation, and so on \cite{b5,b4,b18,b19,b20,b26,d1,d2}.  

In spite of the fact that the particles 
creation effect in external fields was calculated in
numerous papers, there are still some problems  which are interesting to study and
discuss. In the present paper we are going to focus our attention on
the time scenario of the process and to consider it  in
arbitrary dimensions of space-time to be able  analyze its 
dependence on the dimension. To fulfill the first part of the
program we consider special external  fields  which act  effectively
during a finite
time and then compare results  with ones in a constant field. In fact,
such a consideration plays also the role of a
regularization and helps to solve
divergence problems which appear in constant external fields.   
The  dimensional analysis may be interesting in relation with the study of 
multidimensional versions of field theories and gravity. Lower dimensions, e.g.  $2+1$
dimensions can be of a  particular
interest. Field theoretical models
in such  dimensions \cite{b7} attracted in the last years  a great
attention due to various reasons: e.g. nontrivial
topological properties, and especially the possibility of the
existence of particles with fractional spins and exotic statistics
(anyons), having probably applications to fractional Hall effect,
high-$T_c$ superconductivity and so on \cite{b8}.

For calculations we
are using the general approach, which was elaborated in the framework of the field theory
for such kind of problems   
\cite{b9a,b15,b16,b6}. According to this formulation  all the
information about the processes of
particles scattering and creation by an external field (in zeroth order
with respect to the radiative corrections) can be
extracted from special complete sets of exact solutions of the relativistic
wave equations in the external field (IN- and OUT- solutions). A complete collection of
exact solutions of such equations in $3+1$ QED is presented in the book
\cite{b10}, in particular, IN- and OUT-solutions and related
bibliography can be found in \cite{b6}. That is
why in the beginning we analyze and classify  exact solutions of the
Dirac equation in uniform  external electric
fields in arbitrary space-time dimensions. IN- and OUT-solutions are
presented explicitly for T-constant, adiabatic, and constant electric fields.  
Probabilities of particles scattering, pairs creation,
vacuum-to-vacuum probability and mean numbers of particles created are
calculated  in
arbitrary dimensions for  three types of electric fields mentioned above. 
The full consideration in the case of the
T-constant field, which is most
important for the time analysis, has no $3+1$-dimensional analog
and is presented explicitly for the first time. In spite of some of
the formulas in two other cases have $3+1$-dimensional analog, their
d-dimensional generalization appears to be non-trivial. Moreover, some
of these formulas were not presented even in $3+1$-dimensional case,
for example, the total mean numbers of particles created and vacuum-to-vacuum
probability in the adiabatic field. 

Using general expressions obtained for   electric fields, which act
for a finite time, we study the particle creation effect at small and big
times. Thus,  in
particular, we can estimate a stabilization time of the process, 
the time of a pair formation, and give  a 
quasiclassical interpretation of a pair creation. Besides, this
analysis allows one to select and estimate consistently  time
divergences, which appear in the constant fields. Comparing results
obtained in different time configurations of the electric field, we
estimate the role of switching on and off effects in the vacuum
instability. 

We analyze how the effect of the vacuum instability depends on the
space dimensions,  on the possible boundary conditions, and  on a non
trivial topology.

We consider the possibility to add an 
uniform magnetic field to the electric one and calculate the effect. It turns out that one
can formulate universal rules to generalize all the formulas obtained  in the pure
electric field to the case when the magnetic field is included as
well. Its influence on the vacuum instability is studied.        

Finally, it is shown, taking into account the vacuum level shift, that
the distributions of particles created by the quasiconstant electric
fields can be written in a form, which has a thermal
character and seams to be universal, i.e. is valid for  any theory with
a quasiconstant external fields. Its application, for example, to the
particles creation  in external constant gravitational field, reproduces
 the Hawking temperature exactly.

\section{General consideration in an uniform electric field}

The $d$-dimensional Dirac equation in an external electromagnetic field with
potentials $A_{\mu}(x)$ has the form  (further $\hbar=c=1$)
\begin{equation}\label{e1}
\left(P_{\mu}\gamma^{\mu}-m\right)\psi(x)=0\; ,\;\;\; P_{\mu}=
i\partial_{\mu}-eA_{\mu}(x)\;,
\end{equation}
where $\psi(x)$ is a $2^{[\frac{d}{2}]}$-component column,
$\gamma^{\mu}$ are $\gamma$-matrices in $d$ dimensions \cite{b11},
$$[\gamma^{\mu},\gamma^{\nu}]_{+}=
2\eta^{\mu \nu},\;\;\eta^{\mu
\nu}={\rm diag}(\underbrace{1,-1,-1,\ldots }_{d}),\;\;d=D+1,$$
and $  x=(x^{\mu})=
(x^0,{\bf x}), \;{\bf x}=(x^i),\;
\;\mu=0,1,\ldots,D,\;\;i=1,\ldots,D$.

As usual, it is convenient to present $\psi (x)$ in the form 
\begin{equation}\label{e2}
\psi (x)=\left(P_{\mu}\gamma^{\mu}+m\right)\phi (x)\;.
\end{equation}
Then the functions $\phi$ have to obey the squared Dirac equation in
$d$ dimensions,
\begin{equation}\label{e3}
\left(P^2-m^2 -\frac{e}{2}\sigma^{\mu\nu} F_{\mu\nu} \right)\phi
(x)=0\;,\;\;\;
F_{\mu\nu}=\partial_{\mu} A_{\nu} -\partial_{\nu} A_{\mu}\;,
\;\sigma^{\mu\nu}=\frac{i}{2}[\gamma^{\mu},\gamma^{\nu}]\;.
\end{equation}

Let us consider the field $F_{\mu\nu}$ with
only one nonzero   invariant $I=\frac{1}{2}F_{\mu\nu}F^{\mu\nu}$,
which supposes to be negative, $I< 0$. In this case there exists a
reference frame where only the components $F_{0i}$ of the field 
differ from zero.   That corresponds to a pure electric field, 
which  is a particular case of external fields, violating the vacuum
stability (creating particles). Let this electric field  be 
 uniform. It can be nonstationary, but with a constant direction in
the space. Then one can always direct it  along the axis $x^D$. Thus, 
\begin{equation}\label{e4a}
F_{0i}=(0,\ldots,0,E(x^0)),\;  F_{ik}=0.
\end{equation}
For such a field   we will use the following 
potentials: $A_0=A_1=\ldots=A_{D-1}=0,\;\;A_D=A_D(x^0).$ The constant 
uniform electric field is of a special interest, because
QED with such external field (as with any free external field) can be
considered as  exact QED (without external fields) with some
special initial states of the electromagnetic field
\cite{b11a,b11b,b6}, which provide the corresponding nonzero mean values of the
electromagnetic field. Sometimes an 
alternating electric field can also be treated as a slight nonuniform free field,
which is stipulated by some specific external conditions: existence of
a waveguide \cite{b13}, interference of two coherent waves \cite{b14}
and so on. However, the study of the constant field shows the
existence of
divergences related to the infinite action time  of the field. More
correct consideration demands a regularization in time. For instance, one can
consider a field, which acts only a finite time T, being constant
within this interval. Such an approach allows also to 
avoid problems with the definition of IN- and OUT-states in non-switching
external fields at $x^0\rightarrow \pm\infty$. Another possibility is
to consider an alternating field, which switches on and off adiabatically
at $x^0\rightarrow \pm\infty$, and is quasiconstant at finite
times. In the next section we are going to consider all the
possibilities mentioned to study the time scenario of the particles creation.

Solutions of  the  equation (\ref{e3}) in the field (\ref{e4a}) can be
written in the form
\begin{equation}\label{e4}
\phi_{{\bf p},s,r}(x)=\phi_{{\bf
p},s}(x^0)\exp\{i{\bf p}{\bf x}\}v_{s,\{r\}}, \;\;
r=(r_1,\ldots
,r_{[\frac{d}{2}]-1}),\;s=\pm 1,\;\;r_j=\pm1,
\end{equation}
where $v_{s,\{r\}}$ are some constant
orthonormal spinors,
$v^{\dagger}_{s,\{r\}}v_{s,\{r'\}}=\delta_{r,r'}.$  The eq.(\ref{e3}) allows one to
subject these spinors to some supplementary  conditions,
\begin{eqnarray}\label{e5a}
&&S_{\pm}v_{\mp 1,\{r\}}=0,\;
S_{\pm} =\frac{1}{2}(1\pm\gamma^0\gamma^D),\; {\rm
rank}\;S_{\pm}=J_{(d)}=2^{[\frac{d}{2}]-1}; \\
&&R_{\pm}v_{s,\{\mp 1,\bar{r}\}}=0,\;\;R_{\pm}=
\frac{1}{2}(1\pm \frac{i{\bf \gamma p}_{\perp}}{|{\bf p}_{\perp}|}
),\;\;
\mbox{ rank}R_{\pm}=\frac{1}{2}J_{(d)},\;\;\mbox{if} \;d>3,\nonumber \\
&&\bar{r}=(r_2,\ldots
,r_{[\frac{d}{2}]-1}),\;\;
p^a_{\bot}=p^a,\;a=1,\ldots,D-1,
\;p^D_{\bot}=0\;.\nonumber
\end{eqnarray}
If $d\leq 3$ the quantum numbers $r$ do not appear and for $d=2$ 
the perpendicular components of the momenta are absent.

Taking into account the conditions (\ref{e5a}), one can write an 
equation for the functions $\phi_{{\bf
p},s}(x^0)$, 
\begin{equation}\label{e5}
\left[\frac{d^2}{dx_0^2} + (p_{D} -eA_D (x^0 ))^{2}+{\bf p}_{\perp}^{2} +m^2
+iseE(x^0)\right]
\phi _{{\bf p},s}(x^0)=0\;.
 \end{equation}
A formal transition to the spinless case, which corresponds to the use
of the Klein-Gordon equation instead of the Dirac one, 
can be done by putting $s=0$ in (\ref{e5})   and 
$v_{s,\{r\}}=1$ in (\ref{e4}).

The eq.(\ref{e5}) has two independent
solutions at fixed ${\bf p}$ and $s$. Thus, an additional quantum 
number $\zeta$ appears, $\zeta
=\pm \;.$ Combining two independent solutions, which correspond to
different $\zeta \; $, one
can construct two complete sets of solutions ${}_{\zeta}\phi _{{\bf
p},s}(x^0)$
 and ${}^{\zeta}\phi _{{\bf
p},s}(x^0)$,
obeying the following conditions
\begin{eqnarray}\label{e7a}
&&i\frac{d}{dx^0}\;{}_{\zeta}\phi _{{\bf
p},s}(x^0)={}_{\zeta}{\cal E}_{\bf p}  
\;{}_{\zeta}\phi _{{\bf
p},s}(x^0),\;\;{\rm sign}\;{}_{\zeta}{\cal E}_{\bf p}=
\zeta, \;\;x^0\rightarrow -\infty \; ,\nonumber\\
&&i\frac{d}{dx^0}\;{}^{\zeta}\phi _{{\bf
p},s}(x^0)={}^{\zeta}{\cal E}_{\bf p}  
\;{}^{\zeta}\phi _{{\bf
p},s}(x^0),\;\;{\rm sign}\;{}^{\zeta}{\cal E}_{\bf p}=\zeta,
\;\;x^0\rightarrow +\infty \;.
\end{eqnarray}
They provide in turn the following  behavior 
\begin{eqnarray}\label{e8}
&&H_{o.p.}(x^0)\;{}_{\zeta}\psi_{{\bf p},s,r}(x)={}_{\zeta}{\cal E}_{\bf p}  
\;{}_{\zeta}\psi_{{\bf p},s,r}(x),\;  
\;\;,{\rm sign}\;{}_{\zeta}{\cal E}_{\bf p}=
\zeta, \;x^0\rightarrow -\infty \;, \nonumber \\
&&H_{o.p.}(x^0)\;{}^{\zeta}\psi_{{\bf p},s,r}(x)={}^{\zeta}{\cal E}_{\bf p}  
\;{}^{\zeta}\psi_{{\bf p},s,r}(x),\; {\rm sign}\;{}^{\zeta}{\cal E}_{\bf p}=\zeta,
\;\;x^0\rightarrow +\infty \;, 
\end{eqnarray}
of the corresponding Dirac equation solutions,
 ${}_{\zeta}\psi_{{\bf p},s,r}(x)=(\gamma P+m)\;{}_{\zeta}\phi_{{\bf p},
s,r}(x)$ and ${}^{\zeta}\psi_{{\bf p},s,r}(x)=(\gamma P+m)\;{}^{\zeta}\phi_{{\bf p},
s,r}(x).$ In the eq. (\ref{e8}) $H_{o.p.}=\gamma^0(m+{\bf \gamma P})$ is  
one-particle Dirac Hamiltonian, and ${\cal E}$  are
quasi-energies. The solutions ${}_{\pm}\psi_{{\bf
p},s,r}(x)$ describe particle (+) and antiparticle (-) in the initial
time-instant whereas  ${}^{\pm}\psi_{{\bf p},s,r}(x)$ describe particle (+)
and antiparticle (-) in the final
time-instant\cite{b15,b6}. 

One can see that the solutions with different
$s$ and fixed $\zeta,\;{\bf p},\;r$ are dependent. For example, 
\begin{equation}\label{e6a}
{}_{\zeta}\psi_{{\bf
p},s,r}(x)=\frac{m+ib_{{\bf p},r}}{{}_{\zeta}a_{{\bf p},-s}}\;
{}_{\zeta}\psi_{{\bf p},-s,r}(x),
\end{equation} 
where  $b_{{\bf p},r}=r_1|{\bf p}_{\perp}|$ if $d>3$, $ b_{{\bf
p},r}=p_1$ if $d=3$,
  $b_{{\bf p},r}=0$ if
$d=2$, and ${}_{\zeta}a_{{\bf p},s}$ are some coefficients. 
To see how (\ref{e6a}) appears one can use (\ref{e5a}), (\ref{e5}), (\ref{e7a}) and   
the following consequence of two latter, 
$$ \left[i\frac{d}{dx^0}+s(p_D-eA_D(x^0))\right]{}_{\zeta}\phi_{{\bf
p},s}(x^0)={}_{\zeta}a_{{\bf p},s}\;
{}_{\zeta}\phi_{{\bf p},-s}(x^0)\;,\;\;{}_{\zeta}a_{{\bf
p},-1}\;{}_{\zeta}a_{{\bf p},+1}=
m^2+{\bf p}_{\perp}^2\;. $$
Similar relation holds for ${}^{\zeta}\psi_{{\bf
p},s,r}(x)$. The eq.(\ref{e6a}) means, in fact, that the spin
projections of a particle (+) and an antiparticle (-) can take on only
 $J_{(d)}$ values. Taking that into account, one can only use 
the following independent solutions,
\begin{equation}\label{e6}
{}_{\pm}\psi_{{\bf p},r}(x)=(\gamma P+m){}_{\pm}\phi_{{\bf p},
\pm 1,r}(x)\;,\;\;\;
{}^{\pm}\psi_{{\bf p},r}(x)=(\gamma P+m){}^{\pm}\phi_{{\bf p},\mp 1,r}(x)\;.
\end{equation}

Further, we are going to calculate different matrix elements between the
solutions (\ref{e6}) by means of the conventional time independent Dirac scalar
product
$(\psi ,\psi ')=\int \bar{\psi}(x)\gamma^0\psi ' (x)d{\bf x}$. In the
case under consideration, due to the above mentioned properties (\ref{e5a}),(\ref{e5}) of 
the spinors ${}_{\zeta}\phi_{{\bf p},s,r}(x)$ and ${}^{\zeta}\phi_{{\bf p},s,r}(x)$, 
the scalar product can be reduced to a form which is  convenient
for calculation and,  in particular, does not contain $\gamma$
matrices at all,
\begin{eqnarray}\label{e7}
&&({}^{-}_{+}\psi_{{\bf p},r},{}^{-}_{+}\psi_{{\bf
p}',r'})=i(2\pi )^D
\delta_{r,r'}\delta({\bf p}-{\bf p}')  
\;{}^{-}_{+}\phi^*_{{\bf p},+
1}(x^0) \stackrel{\longleftrightarrow}{\partial_{0}}\left(i\partial_{0}+p_D-eA_D(x^0)\right)
{}^{-}_{+}\phi_{{\bf p},+ 1}(x^0),\;\nonumber \\
&&({}^{+}_{-}\psi_{{\bf p},r},{}^{+}_{-}\psi_{{\bf
p}',r'})=i(2\pi )^D
\delta_{r,r'}\delta({\bf p}-{\bf p}')  
\;{}^{+}_{-}\phi^*_{{\bf p},-
1}(x^0) \stackrel{\longleftrightarrow}{\partial_{0}}\left(i\partial_{0}-p_D+eA_D(x^0)\right)
{}^{+}_{-}\phi_{{\bf p},- 1}(x^0),\;\nonumber \\
&&({}^{-}_{+}\psi_{{\bf p},r},{}^{+}_{-}\psi_{{\bf p}',r'})=i(2\pi)^D
\delta_{r,r'}\delta({\bf p}-{\bf p}')
(m-ib_{{\bf p},r}) \;{}^{-}_{+}\phi^*_{{\bf p},+1}(x^0)
\stackrel{\longleftrightarrow}
{\partial_{0}}
{}^{+}_{-}\phi_{{\bf p},-1}(x^0),  
\end{eqnarray}
where
$\stackrel{\longleftrightarrow}{\partial_{0}}=
\stackrel{\longrightarrow}{\partial_{0}}-\stackrel{\longleftarrow}{\partial_{0}}.$
(The right side of (\ref{e7}) reproduces the corresponding 
Klein-Gordon scalar product if  one puts formally 
 $\zeta [i\partial_{0}\pm (p_D-eA_D(x^0))]=m-ib=1$.)
 
One can see from (\ref{e7a}) and (\ref{e7}) that the solutions
(\ref{e6})
can be normalized to obey the orthonormality relations, 
\begin{equation}\label{e9}
\left( {}_{\zeta}\psi_{{\bf p},r},\;{}_{\zeta '}\psi_{{\bf
p}',r'}\right)=\delta_{\zeta,\zeta '}\delta_{r,r'}\delta({\bf p}-{\bf p}'),\;\;
\left( {}^{\zeta}\psi_{{\bf p},r}\;{}^{\zeta '}\psi_{{\bf
p}',r'}\right)=\delta_{\zeta,\zeta '}\delta_{r,r'}\delta({\bf p}-{\bf p}').
\end{equation}
Moreover, each set of solutions ${}_{\zeta}\psi_{{\bf p},r}(x)$ and
 ${}^{\zeta}\psi_{{\bf p},r}(x)$ forms a complete system, thus, we are
dealing with the so-called IN-  and
OUT-sets of solutions correspondingly \cite{b15,b16,b6}.

Using  (\ref{e7}),  one can find   decomposition coefficients
$G\left({}_{\zeta}|{}^{\zeta '}\right)$ of the OUT-solutions in the IN-solutions,
\begin{equation}\label{e10}
{}^{\zeta}\psi(x)={}_{+}\psi(x)G\left({}_{+}|{}^{\zeta}\right)+{}_{-}\psi(x)
G\left({}_{-}|{}^{\zeta}\right)\;.
\end{equation}
The matrices $G\left({}_{\zeta}|{}^{\zeta'}\right)$ obey the
following relations,
\begin{eqnarray}\label{e10a}
&&G\left({}_{\zeta}|{}^{+}\right)G\left({}_{\zeta}|{}^{+}\right)^{\dagger}+\kappa 
G\left({}_{\zeta}|{}^{-}\right)G\left({}_{\zeta}|{}^{-}\right)^{\dagger}=
(\zeta {\bf I})^{\frac{1-\kappa}{2}},\nonumber \\                  
&&G\left({}_{+}|{}^{+}\right)G\left({}_{+}|{}^{+}\right)^{\dagger}+\kappa
G\left({}_{+}|{}^{-}\right)G\left({}_{-}|{}^{-}\right)^{\dagger}=0\;,
\end{eqnarray}
where ${\bf I}$ is the unit matrix and  $\kappa=\pm 1$ for fermions
and  bosons respectively. Relations
(\ref{e10a}) can be derived from
the conditions  (\ref{e9}). Due to eq.(\ref{e7}) we can easily see
that the  matrices
$G\left({}_{\zeta}|{}^{\zeta'}\right)$
are diagonal,
\begin{equation}\label{e37a}
G\left({}_{\zeta}|{}^{\zeta'}\right)_{{\bf
p} ,r,{\bf p}',r'}=\delta_{r,r'}
\delta({\bf p}-{\bf
p}')\;g\left({}_{\zeta}|{}^{\zeta'}\right).
\end{equation}
All the information about the  processes   of particles creation, annihilation, and
scattering in an external field (without radiative corrections) one
can extract from the matrices   
$G\left({}_{\zeta}|{}^{\zeta'}\right)$  because they define a  
canonical transformation between IN and OUT creation and annihilation
operators in the generalized Furry representation \cite{b15,b16,b6},
\begin{eqnarray}\label{e10b}
&&a^{\dagger}(out)=a^{\dagger}(in)G\left({}_{+}|{}^{+}\right)+
b(in)G\left({}_{-}|{}^{+}\right),\nonumber\\
&&b(out)=a^{\dagger}(in)G\left({}_{+}|{}^{-}\right)+
b(in)G\left({}_{-}|{}^{-}\right).
\end{eqnarray}
 Here 
$a_{n}^{\dagger}(in)$, $b_{n}^{\dagger}(in)$, $a_{n}(in)$, $b_{n}(in)$
 are  creation and annihilation operators of IN-particles and
antiparticles respectively  
 and $a_{n}^{\dagger}(out)$,$b_{n}^{\dagger}(out)$,
 $a_{n}(out),\;b_{n}(out)$ are ones of OUT-particles and
antiparticles, $n$ presents momentum ${\bf p}$ and spin projections $r$. For example, 
the mean numbers  of  particles  created
(which are also equal to the numbers of pairs created) 
by the external field from the IN-vacuum $|0,in>$  with a given 
momentum ${\bf p}$ and spin projections $r$ is
\begin{equation}\label{e11}
 N_{{\bf p},r}= <0,in|a_{{\bf p},r}^{\dagger}(out)a_{{\bf p},r}(out)|0,in>=
\left|g\left({}_{-}|{}^{+}\right)
\right|^2.
\end{equation} 
Here  the standard 
 volume regularization was used, so that $\delta ({\bf p}-{\bf
p}')\rightarrow \delta_{{\bf p},{\bf p}'} $.
The probabilities  of  a particle scattering and of a pair creation
have the following  forms respectively
\begin{eqnarray}
&& P(+|+)_{{\bf p},r,{\bf p'},r'}
=|<0,out|a_{{\bf p},r}(out)a_{{\bf p'},r'}^{\dagger}(in)|0,in>|^2=\delta_{r,r'}
\delta_{{\bf p},{\bf
p}'}
\frac{1}{1-\kappa N_{{\bf
p},r}}P_v\;,\label{e12a} \\
&&P(-+|0)_{{\bf p},r,{\bf p'},r'}=|<0,out|b_{{\bf p},r}(out)a_{{\bf p'},r'}(out)|0,in>|^2=
\delta_{r,r'}
\delta_{{\bf p},{\bf
p}'}
\frac{N_{{\bf p},r}}{1-\kappa N_{{\bf
p},r}}P_v\;,\label{e12b}
\end{eqnarray}
where $|0,out>$ is the OUT vacuum and 
\begin{equation}\label{e13}
P_v=|<0,out|0,in>|^2  =\exp\left\{\kappa \sum_{{\bf p},r}\ln\left(
1-\kappa N_{{\bf
p},r}\right) \right\}\;,\label{e39ab}
\end{equation}
is the probability  for a vacuum to remain a vacuum.
The probabilities  for an antiparticle scattering and a pair
annihilation  are described by the same expressions $P(+|+)$ and $P(-+|0)$ respectively.

Thus, to be able to  calculate   the quantities
(\ref{e11})-(\ref{e13}), in the case under consideration, one has to 
find  solutions of the ordinary differential equation
(\ref{e5}), which is in fact Schr\"odinger equation for a linear
oscillator with time-dependent frequency. However, one can make some
general conclusions, which do not depend on
the concrete time dependence of the electric field in eq.(\ref{e5}).
 First of all, the matrices $G\left({}_{\zeta}|{}^{\zeta '}\right)$
are diagonal in all the quantum numbers introduced. Second, the
quantum numbers $r$ do not enter in the eq.(\ref{e5}) and,  due to
the structure of the scalar product (\ref{e7}), the matrices can depend
on $r$ via a phase only. That is why all the  probabilities and the
mean numbers  do not depend on $r$, so that in the fermionic
case the total (summed over all $r$) 
probabilities and the mean numbers   are $J_{(d)}$ times greater than the
correspondent differential
quantities. For example, the total number of  particles  created with a
given momentum ${\bf p}$ is
\begin{equation}\label{e1a}
N_{\bf p}=\sum_{r}N_{{\bf p},r}=
J_{(d)}N_{{\bf p},r}.
\end{equation}
 Finally, it is clear that due to the structure of the eq. (\ref{e5}) and
the scalar product (\ref{e7}) the dimensionality $d$
enter in the differential probabilities and mean values via the
combination ${\bf p}^2_{\perp}$ only.

\section{T-constant, adiabatic, and constant electric fields }

\subsection{ $T$-constant field}

To  analyze the time dependence of the particles creation effects let
us consider the field (\ref{e4a}) with $E(x^0)$ having the form
\begin{equation}\label{e31a}
E(x^0)=\left\{\begin{array}{ll}
         0,\;\;&x^0\in I\\
         E,\;\;&x^0\in II\\
         0,\;\;&x^0\in III\;,
\end{array}\right. 
\end{equation}
where the time intervals  are: $I=(-\infty
,t_1),$ $II=[t_1 ,t_2],$ $III=(t_2,+\infty),$ $t_2-t_1=T,\;t_2=-t_1,$
and $eE>0$ is chosen. 
 Thus, in fact, we  consider a constant electric field $E$, which is acting 
  a finite time T. Further we will call it $T$-constant field.
The corresponding potential $A_D(x^0)$ can be chosen in the form
\begin{equation}\label{e32a}
A_D(x^0)=\left\{\begin{array}{ll}
         Et_1,\;\;&x^0\in I\\
         Ex^0,\;\;&x^0\in II\\
         Et_2,\;\;&x^0\in III\;.
\end{array}\right. 
\end{equation}
In each interval $I,\;II,\;III$ the equation (\ref{e5}) has two independent
solutions, which are correspondingly in $I$: $\exp\{-ip_0(t_1)x^0\}$ and
$\exp\{+ip_0(t_1)x^0\}$, in  $II$: 
 $D_{\nu -\frac{1+s}{2}}\left[ (1-i)\xi\right]$ and
 $D_{-\nu-\frac{1-s}{2}}\left[ (1+i)\xi\right]$, and 
 in  $III$: 
$\exp\{-ip_0(t_2)x^0\}$ and $
\exp\{+ip_0(t_2)x^0\}$, 
where $D_{\nu}(z)$ are  
Weber parabolic cylinder functions (WPC-functions) 
\cite{b12}, and
\[
\nu=\frac{i\lambda}{2},\;\;\lambda=\frac{m^2+{\bf p}^2_{\bot}}{eE},\;\;
\xi (x^0)=\frac{eEx^0-p_D}{\sqrt{eE}},\;\;
p_0(x^0)=\sqrt{m^2+{\bf p}^2_{\perp}+(p_D-eA_D(x^0))^2}\;\;.
\]
Using them and conditions (\ref{e7a}), one can construct IN- and
OUT-solutions ${}_{-}\psi_{{\bf p},r}(x)$ and $^{+}\psi_{{\bf
p},r}(x)$  (see Sect.II). The corresponding expressions
 for ${} _{-}\phi_{{\bf
 p},-1}(x^0)$ and ${} ^{+}\phi_{{\bf
 p},-1}(x^0)$ are of the form  

\begin{eqnarray}\label{e33a}
&&{} _{-}\phi_{{\bf
 p},-1}(x^0)  \\
&&=C_1\left\{\begin{array}{ll}
\exp\{+ip_0(t_1)(x^0-t_1)\}, \;\;&I\\
a_1D_{\nu }\left[ (1-i)\xi\right]+
a_2 D_{-\nu-1}\left[ (1+i)\xi\right]        
 ,\;\;&II\\
g(^+|_-) \exp\{-ip_0(t_2)(x^0-t_2)\}+g(^-|_-)
\exp\{+ip_0(t_2)(x^0-t_2)\}
  ,\;\;&III\;;
\end{array}\right. \nonumber
\end{eqnarray}
\begin{eqnarray}\label{e34a}
&&{} ^{+}\phi_{{\bf
 p},-1}(x^0) \\
&&=C_2\left\{\begin{array}{ll}
g(_+|^+) \exp\{-ip_0(t_1)(x^0-t_1)\}+g(_-|^+)
\exp\{+ip_0(t_1)(x^0-t_1)\}, \;\;&I\\
a'_1D_{\nu }\left[ (1-i)\xi\right]+
a'_2 D_{-\nu-1}\left[ (1+i)\xi\right]        
 ,\;\;&II\\
\exp\{-ip_0(t_2)(x^0-t_2)\}\;,  \;\;&III\;,
\end{array}\right.\nonumber 
\end{eqnarray}
where the normalization constants are $$ C_{i}=(2\pi
)^{-D/2}(2p_0(t_i)q_i)^{-1/2},\; q_i=p_0(t_i)-(-1)^i(p_D-eA_D(t_i)),\;i=1,2.$$
To provide the continuity of the solutions in the time instants $t_1$
and $t_2$ one has to impose the following conditions
$${}_- ^{+}\phi_{{\bf p},-1}(t_i+0)={}_- ^{+}\phi_{{\bf p},-1}(t_i-0),\;\;
  \frac{d}{dx^0}{}_- ^{+}\phi_{{\bf p},-1}(t_i+0)=
\frac{d}{dx^0}{}_- ^{+}\phi_{{\bf p},-1}(t_i-0)\;,$$
which allow one to define step by step all the coefficients $a_i,
\;a'_i,$ and
$g\left({}_{\pm}|{}^{+}\right)$,   $g\left({}^{\pm}|{}_{-}\right).$
The first ones are  
$$a_i=-(-1)^i\frac{p_0(t_1)f^{(+)}_i(t_1)}{M\sqrt{2eE}}\;,\;\;
a'_i=(-1)^i\frac{p_0(t_2)f^{(-)}_i(t_2)}{M\sqrt{2eE}}\;,$$
 where
$$M=D_{\nu }( z)\;\frac{d}{dz}D_{-\nu-1}( iz)-D_{-\nu
-1}(iz)\frac{d}{dz}D_{\nu}(z)=\exp\{-(\nu +1)\frac{i\pi} {2}\}$$
is the Wronskian determinant \cite{b12}, and
$$f^{(\pm)}_1(x^0)=\left(1\pm\frac{i\partial_0}{p_0(x^0)}\right)
D_{-\nu-1}\left[ (1+i)\xi\right],\;\;
f^{(\pm)}_2(x^0)=\left(1\pm\frac{i\partial_0}{p_0(x^0)}\right)
D_{\nu}\left[ (1-i)\xi\right].$$
They can be used to define the latter coefficients. From those we need
to know explicitly only 
  $g\left({}_{-}|{}^{+}\right)$ and
$g\left({}^{+}|{}_{-}\right)$, which are
\begin{eqnarray}\label{e37b}
&&g\left({}_{-}|{}^{+}\right)=
\exp\{(\nu +1)\frac{i\pi} {2}\}\left(\frac{p_0(t_1)q_1p_0(t_2)}{8eEq_2}\right)^{\frac{1}{2}}
\left[ f^{(-)}_2(t_2) f^{(-)}_1(t_1) - f^{(-)}_1(t_2) f^{(-)}_2(t_1)
\right],\nonumber\\
&&g\left({}^{+}|{}_{-}\right)=
\exp\{(\nu +1)\frac{i\pi} {2}\}\left(\frac{p_0(t_1)q_2p_0(t_2)}{8eEq_1}\right)^{\frac{1}{2}}
\left[ f^{(+)}_2(t_2) f^{(+)}_1(t_1) - f^{(+)}_1(t_2) f^{(+)}_2(t_1)
\right].
\end{eqnarray}
One can see that the coefficients (\ref{e37b}) obey the properties 
\begin{equation}\label{e37bb}
\left.
g\left({}^{+}|{}_{-}\right)\right|_{p_D\rightarrow
-p_D}=-g\left({}_{-}|{}^{+}\right),\;\;g\left({}^{+}|{}_{-}\right)=g\left({}_{-}|{}^{+}\right)^*.
\end{equation}
The first one can be verified directly, whereas the second one  is
easy to derive comparing representations of the scalar
product (\ref{e7}) in the time instants $t_1$ and $t_2$. Thus, one can conclude that 
$\left|g\left({}_{-}|{}^{+}\right)\right|$ is an even function of the 
momentum $p_D$.

To calculate the probabilities and the mean numbers
according to the formulas (\ref{e11}-\ref{e13}) we  need really 
to know only the coefficients $g\left({}_{-}|{}^{+}\right)$. 
Comparing (\ref{e10}) and (\ref{e33a}), (\ref{e34a}), we conclude that
the relations (\ref{e37a}) hold and we have the expression for the mean
numbers of pairs created in the form (\ref{e11}), in which $g\left({}_{-}|{}^{+}\right)$ 
is defined by (\ref{e37b}). In 
fact, the mean numbers $N_{{\bf p},r}$ define all the probabilities via
the formulas (\ref{e12a}-\ref{e13}).  
As it was shown above this function is even in all the momenta ${\bf p}$,
including $p_D$ and does not depend on the spin quantum number
$r$. Using the recipe presented in Sect.II,  it is easy to get an explicit form
$N_{\bf p}$ for the the bosonic case from the fermionic one  (\ref{e11}) and (\ref{e37b}).

Now we are going to analyze the dependence of all the characteristics
on the time $T$ and on the momenta. 
One has to remark that the dependence of the longitudinal momentum
$p_D$ is of a special interest. This dependence is essentially
correlated with the  $T$-dependence.
One can see, e.g. from the
four-dimensional  case \cite{b9a,b6}, that in the constant field 
($T= \infty$)  all the characteristics do not depend on
the momentum $p_D$. This is a source of some kind  divergences  if
one is interested in the total characteristics, which require
integration over $p_D$. Due to the reasons mentioned above  it is enough to analyze only the
quantity $N_{{\bf
p},r}$.  As to the momentum $p_D$, one can restrict itself only by 
 $p_D$ positive or $p_D$ negative.
Remark that the momenta $p_D$ enter in all the formulas  via two
dimensionless parameters $\xi_1$ and $\xi_2$ only,  
$$\xi_1=\xi (-\frac{T}{2})=\frac{1}{\sqrt{eE}}(-eE\frac{T}{2}-p_D),\;
\xi_2=\xi (+\frac{T}{2})=\frac{1}{\sqrt{eE}}(+eE\frac{T}{2}-p_D),$$
which in turn appear in the WPC-functions. Thus, in fact, one needs to
analyze the dependence of the  latter on $\xi$. It is  convenient to consider the region
$-\sqrt{eE}\frac{T}{2}\leq \xi_1<+\infty\;,
\;\xi_2\geq\sqrt{eE}\frac{T}{2}$, which corresponds  to $p_D$ negative, 
$0\leq -p_D<+\infty$, in particular, there
always  $\xi_2>\xi_1$.

In the region $\xi_1\geq K,$ where $K$ is a given
number $K>>1+\lambda $ (in terms of the momentum  this region
corresponds to $|p_D|\geq eE\frac{T}{2}+K\sqrt{eE}$)    
one can use the  asymptotic expansion of  WPC- functions
\cite{b12},
\begin{equation}\label{e40ab}
D_{\nu}(z)=z^{\nu}\exp\left\{-z^2/4\right\}\left(\sum_{n=0}^{N}\frac{(-\frac{1}{2}\nu)_n(\frac{1}{2}-\frac{1}{2}\nu)_n}{n!(-\frac{1}{2}z^2)^n}+
O(|z|^{-2(N+1)})\right),\;
|\arg z|<\frac{3}{4}\pi,
\end{equation}
to conclude that at any $T$ the behavior of the mean numbers (\ref{e11}), (\ref{e37b})  is
\begin{equation}\label{e40abc}
N_{{\bf p},r}=
O\left( \left[\frac{\lambda}{\xi_1^2}\right]^3\right).
\end{equation}
For small $T$, 
 $T<<\frac{1}{\sqrt{eE}}$, and $|p_D|<<eE\frac{T}{2}$, one
can get 
\begin{equation}\label{e41a}
N_{{\bf p},r}=\frac{eET^2}{4\lambda+eET^2}\left[1+O(\sqrt{eE}T)\right].
\end{equation}
At $T<<\frac{\sqrt{\lambda}}{\sqrt{eE}}$ and 
$T<<\frac{1}{\sqrt{eE\lambda}}$ the form (\ref{e41a}) reduces to  
$N_{{\bf p},r}=eET^2/(4\lambda)$ and coincides with one which can be
derived in the frame of perturbation theory with respect to the
external field.

The most important region  for the time divergences is one of big $T$, 
namely, let us consider
$T>>\frac{1}{\sqrt{eE}}(1+\lambda)$. In this case $\xi_2$ is
always big and positive $\xi_2>>1+\lambda$, so that
the asymptotic expansion (\ref{e40ab})  can be used for any given momentum
$p_D$. As to the parameter $\xi_1$,  the whole interval
$-\sqrt{eE}\frac{T}{2}\leq \xi_1<+\infty$ can be divided in three regions:
\[
a) \; -\sqrt{eE}\frac{T}{2}\leq\xi_1\leq -K;\;\;\;
b)  \; -K<\xi_1<K;\;\;\; c)\; \xi_1\geq K\;.
\]
The mean numbers  $N_{{\bf p},r}$ were  estimated in the region c)
before, see  (\ref{e40abc}). In the region a)
one can use some relations between the WPC-functions
\cite{b12}, for example,  $D_{\nu} (z)=\exp\{
i\pi\nu\}D_{\nu}(-z)-\frac{\sqrt{2\pi}}{\Gamma (-\nu )}\exp \frac{
i\pi\nu}{2}D_{-\nu-1}(-iz),$
and the  asymptotic expansion (\ref{e40ab}). Then one  finds
\begin{equation}\label{e40a}
N_{{\bf p},r}=e^{-\pi\lambda}\left[1+O\left(\left[\frac{1+\lambda}{\xi_1}\right]^3\right)+
O\left(\left[\frac{1+\lambda}{\xi_2}\right]^3\right)\right]
,\;\;-\sqrt{eE}\frac{T}{2}\leq\xi_1\leq -K.
\end{equation}

The latter expression allows one to consider the limit $T\rightarrow\infty$ at
any given ${\bf p}$. In this limit the mean numbers take a simple form
\begin{equation}\label{e40ad}
N_{{\bf p},r}=e^{-\pi\lambda}\;.
\end{equation}
Thus, when the electric field is acting for a long enough time, the mean numbers
of particles created in a given quantum state are stabilized and coincide
with expressions which were obtained in the constant electric field in 3+1 QED \cite{b9a}.
(The  stabilization (\ref{e40a})   was first remarked in
\cite{b9ab}  for
particles created with zero momenta, using a finite action electric field in 3+1 QED.)
 One can also estimate a
characteristic time of such a stabilization. To this end one can see
that $\frac{1+\lambda}{\sqrt{eE}T}$ is a small
parameter in the decomposition (\ref{e40a}) in case
  $|p_D|<<eE\frac{T}{2}$. If
$T>>T_0,\;\;T_0=\frac{1+\lambda}{\sqrt{eE}}$ then the mean numbers 
are stabilized and $T_0$ is the above mentioned characteristic  time.
 
The intermediate region b) does not allows one to use
an asymptotic expansion of WPC-functions to analyze the $\xi_1$
dependence. However, one can make some  conclusions about its
contribution in integrals over the momenta. For example, due to the Fermi
statistics $N_{{\bf p},r}$ is
always smaller then unity, that is why the integral over the momentum $p_D$ in the
region b) is less then
$2\sqrt{eE}K$ and is not essential in comparison with the same
integral in the region a) at $T\rightarrow\infty$.

Using these considerations, one can now estimate the sum 
 over the longitudinal momentum $p_D$ of $N_{{\bf
p},r}$, which is the mean number
of particles created with all possible momenta $p_D$.
To do this  we go over to the integral,
$\sum_{p_D}\rightarrow \frac{L}{2\pi}\int dp_D$, where $L$ is the length in the
direction $x^D$.
As was shown above, at $T>>\frac{1}{\sqrt{eE}}(1+\lambda)$,   
 $N_{{\bf p},r}$ is quasiconstant in the area a),  the 
asymptotics in the area c) has the form (\ref{e40abc}) and the contribution
to the integral form the area b) is less then $\sqrt{eE} KL/\pi$.
Then one can conclude
\begin{equation}\label{e42a}
N_{{\bf p_{\perp}},r}=\frac{L}{2\pi}\int_{-\infty}^{+\infty}N_{{\bf
p},r}dp_D=\frac{\sqrt{eE}L}{2\pi}\left[\sqrt{eE}Te^{-\pi\lambda}+O(K)
\right].
\end{equation}
Thus, when $T>>\frac{1}{\sqrt{eE}}K>>\frac{1}{\sqrt{eE}}(1+\lambda)$ 
 we can effectively replace the integral over $p_D$ by $eET$ and write
\begin{equation}\label{e41ab}
N_{{\bf p_{\perp}},r}=\Delta_{long}  e^{-\pi\lambda},\;\;
\Delta_{long}=\frac{1}{2\pi}eELT\;.
\end{equation}
The factor  $\Delta_{long}$ can be can interpreted as the total number of
states with the longitudinal momenta $p_D$ of  particles created.

It turns out that the  expressions for $N_{{\bf p},r}$ and $N_{{\bf
p_{\perp}},r}$ at big $T$ for scalar particles coincide with the  
ones for spinor particles.

To get the total number $N$ of  particles created one can sum  over the  spin
projections, using eq.(\ref{e1a}), 
 and then over  the transversal momenta. The latter sum can be 
 easily transformed into an integral,
\begin{equation}\label{e1b}
N=\sum_{\bf p}N_{\bf p}=\frac{V_{(d-1)}}{(2\pi)^{d-1}}\int d{\bf
p}N_{\bf p},
\end{equation}
where $V_{(d-1)}$ is (d-1)-dimensional spatial volume. Thus, on gets
\begin{equation}\label{e19}
N= J_{(d)}^{\frac{1+\kappa}{2}}\frac{V_{(d-1)}Tm^d}{(2\pi)^{d-1}}\left(\frac{E}{E_c}\right)^
{\frac{d}{2}}\exp\left\{-\pi\frac{E_c}{E}\right\}, \;\;J_{(d)}=2^{[\frac{d}{2}]-1},
\end{equation}
where $E_c=m^2/e$   is the critical field strength. 
As one can see the  velocity of particles creation is constant at big T.
The same result was obtained in 3+1 dimensions in the framework of the
QED at finite temperature, using the functional Schr\"odinger
picture \cite{d3}.

The vacuum-to-vacuum transition probability (\ref{e39ab})
   can be calculated, using both kinds of regularizations, with respect to the
volume and to the time. Thus, we get the d-dimensional analog of the
well-known Schwinger formula \cite{b2},    
\begin{equation}\label{e20}
P_v=\exp\left\{-\mu 
N\right\},\;\; 
\mu=\sum_{n=0}^{\infty}\frac{(-1)^{(1-\kappa)\frac{n}{2}}}{(n+1)^{\frac{d}{2}}}
\exp \left\{-n\pi
\frac{E_c}{E}\right\} \;.
\end{equation}
Returning to the Schwinger result, one ought to say that  it was, in fact, obtained from the
constant field consideration by means of a regularization. Thus, the space-time
volume $VT$  appeared in his formula. 
One ought to mention the paper \cite{d2} where the Schwinger result in
the constant electric field was recovered, using the functional
Schr\"odinger equation.

\subsection{ Adiabatic field}

Let us consider an alternating uniform electric field (\ref{e4a}),
where the function $E(x^0)$ has the  following form
\begin{equation}\label{e2a} 
E(x^0)=E\cosh^{-2}\left(\frac{x^0}{\alpha}\right).
\end{equation}
Such a field  switches on and off adiabatically
at $x^0\rightarrow \pm\infty$, and is quasiconstant at finite
times. We will call it adiabatic field. The corresponding nonzero potential is
\begin{equation}\label{e3a} 
A_D(x^0)=\alpha E\tanh\frac{x^0}{\alpha}.
\end{equation}
In this case solutions of eq.(\ref{e5}) can be written in terms of
hypergeometric functions $F(a,b;c;y)$ \cite{b12}, for example,
\begin{eqnarray}\label{e4aa}
&&{} _{\zeta}\phi_{{\bf
 p},+1}(x^0)={}
_{\zeta}Ce^{-i\omega_-x^0}\left(1+e^{\frac{2x^0}{\alpha}}\right)^{\frac{i\alpha}{2}
(\omega_- -\omega_+)}{} _{\zeta}u(x^0),\\
&&{} _{+}u(x^0)=F(a,b;c;y),\;\;{}
_{-}u(x^0)=y^{1-c}F(a-c+1,b-c+1;2-c;y),\nonumber \\
&&
a=\frac{i\alpha}{2}(2eE\alpha+\omega_+-\omega_-),\;\;
b=1+\frac{i\alpha}{2}(-2eE\alpha+\omega_+-\omega_-),\nonumber\\
&&c=1-i\alpha\omega_-,\;\;y=\frac{1}{2}(1+\tanh\frac{x^0}{\alpha}),\;\;
\omega_{\pm}=\sqrt{m^2+{\bf
p}^2_{\perp}+(p_D\mp eE\alpha)^2}, \nonumber
\end{eqnarray}
where ${}
_{\zeta}C$ are some normalization constants. Considering the
asymptotic of the functions (\ref{e4aa}) at $x^0\rightarrow-\infty$,
(in this case 
$F(a,b;c;y)=1$ \cite{b12}), one can verify that the relations
(\ref{e7a}) hold and 
${}_{\zeta}{\cal E}_{\bf p}=\zeta \omega_-$. Moreover, the solutions
at $x^0\rightarrow-\infty$ describe free particles. By analogy one
can construct  solutions ${} ^{\zeta}\phi_{{\bf
 p},+1}(x^0)$, which   describe free
particles with energies 
${}^{\zeta}{\cal E}_{\bf p}=\zeta \omega_+$ at
$x^0\rightarrow+\infty$. Their asymptotic form  at
$x^0\rightarrow+\infty$ is ${} ^{\zeta}\phi_{{\bf
 p},+1}(x^0)={}
^{\zeta}C\exp(-i\zeta\omega_+x^0)$. To calculate the coefficients  
$G\left({}_{+}|{}^{-}\right)$ from (\ref{e10}) it is enough to know
the corresponding asymptotic
${} _{+}\phi_{{\bf
 p},+1}(x^0)$ and ${} ^{-}\phi_{{\bf
 p},+1}(x^0)$,  let say at $x^0\rightarrow+\infty$,  and normalization constants ${}
^{-}C=(2\pi)^{-D/2}\left(2\omega_+(\omega_+-p_D+eE\alpha)\right)^{-1/2},\;\;
{}
_{+}C=(2\pi)^{-D/2}\left(2\omega_-(\omega_-
+p_D+eE\alpha)\right)^{-1/2}$. Thus, we get the mean numbers of fermions created, 
\begin{equation}\label{e18a}
N_{{\bf p},r}=\frac{\sinh\left[ \frac{\pi\alpha}{2}
(2eE\alpha+\omega_- -\omega_+
)\right] \;
\sinh\left[ \frac{\pi\alpha}{2}(2eE\alpha+\omega_+ -
\omega_-)\right]
}{\sinh (\pi\alpha\omega_+ )\sinh (\pi\alpha \omega_-)}\;.
\end{equation}
In 3+1 QED the corresponding formula was found first in \cite{b9b}.
For scalar particles it has a different form 
\begin{equation}\label{e18b}
N_{{\bf
p},r}=\frac{\cosh^2\left[\pi\sqrt{(eE\alpha^2)^2-\frac{1}{4}}\right]+
\sinh^2\left[ \frac{\pi\alpha}{2}(\omega_+-
\omega_-)\right]}
{\sinh (\pi\alpha \omega_+)\sinh (\pi\alpha \omega_-)}\;.
\end{equation}

Let us consider the $\alpha$-dependence of these expressions. For small 
$\alpha$, $\alpha<<\frac{1}{eE}\sqrt{m^2+{\bf p}^2}$,
 when the potential  changes sharply, we get for fermions
\begin{equation}\label{e18aa}
N_{{\bf
p},r}=
\frac{(\pi eE\alpha^2)^2\left(1-\frac{p_D^2}{m^2+{\bf
p}^2}\right)}
{\sinh^2 (\pi\alpha \sqrt{m^2+{\bf p}^2)} }\;,
\end{equation}
and for bosons
\begin{equation}\label{e18ab}
N_{{\bf
p},r}=
\frac{(\pi eE\alpha^2)^2\left[(eE\alpha^2)^2+\frac{p_D^2}{m^2+{\bf
p}^2}\right]}
{\sinh^2 (\pi\alpha \sqrt{m^2+{\bf p}^2)} }\;.
\end{equation}
Small $\alpha$ in the case under consideration corresponds in a sense to small $T$
of the T-constant field. Thus, we have to compare the expressions
(\ref{e41a}) and (\ref{e18aa}). One can see that they are quite
different, so that the effects of switching on and off are essential
at small times.

Further let us consider big $\alpha$ only, $\alpha>>\frac{1}{\sqrt{eE}}(1+\sqrt{\lambda})$.
Then the  mean numbers  for fermions  and bosons have the
same form,
\begin{equation}\label{e19a}
N_{{\bf
p},r}=\exp\left\{-\pi\alpha (\omega_+ +\omega_- -2eE\alpha
)\right\}.
\end{equation}
Let us take small longitudinal momenta
$|p_D|<<eE\alpha$, then
\begin{equation}\label{e19b}
N_{{\bf p},r}=
\exp\left\{-\pi\lambda\left[1+\left(\frac{p_D}{eE\alpha}\right)^2\right]\right\}.
\end{equation}
Considering the limit $\alpha\rightarrow\infty$, one gets the formula
(\ref{e40ad}). That means that the effects of switching on and off are
not essential at big times and small longitudinal momenta. 
For big longitudinal momenta $|p_D|>>eE\alpha$, 
the mean numbers of particles created  are exponentially
small,
\begin{equation}\label{e19ba}
N_{{\bf p},r}=
\exp\left\{-2\pi\alpha(|p_D|-eE\alpha)\right\}.
\end{equation}
Let us find the total numbers of particles created with all the
longitudinal momenta at any fixed ${\bf p}_{\perp},\;r$.  Passing from
the summation over $p_D$  to the corresponding integration, we get
\begin{equation}\label{e20a}
N_{{\bf p}_{\perp},r}=
\frac{LeE\alpha}{2\pi\sqrt{\lambda}}
e^{-\pi\lambda}\;.
\end{equation}
Comparison with the formula (\ref{e41ab}) shows that
the adiabatic field at big times (big $\alpha,\;\alpha >>
\frac{1}{\sqrt{eE}}(1+\sqrt{\lambda}$)
and fixed ${\bf p}_{\perp},\;r$ is equivalent to the $T$-constant field 
at $T>>\frac{1}{\sqrt{eE}}(1+\lambda)$ with the identification
$\alpha=\sqrt{\lambda}T$. To do summation over all 
transversal momenta, it is convenient to use the
representation
$$\frac{1}{\sqrt{\lambda}}=2\int_{0}^{\infty}\exp(-\pi\lambda s^2)ds.$$
Then the total number $N$ of the particles created reads
\begin{equation}\label{e19c}
N= J_{(d)}^{\frac{1+\kappa}{2}}\frac{V_{(d-1)}\alpha\delta
m^d}{(2\pi)^{d-1}}\left(\frac{E}{E_c}\right)^
{\frac{d}{2}}\exp\left\{-\pi\frac{E_c}{E}\right\}\;,
\end{equation}
where
$$\delta=\int_{0}^{\infty}dtt^{-\frac{1}{2}}(t+1)^{-\frac{d-2}{2}}\exp(-t\pi\frac{E_c}{E})=
\sqrt{\pi}\Psi\left(\frac{1}{2},-\frac{d-2}{2};\pi\frac{E_c}{E}\right)$$
is expressed via the  confluent hypergeometric function \cite{b12}.  The
vacuum-to-vacuum transition probability $P_v$ has the form
\begin{eqnarray}\label{e20c}
&&P_v=\exp\left\{-\mu 
N\right\},\;\; 
\mu=\sum_{n=0}^{\infty}\frac{(-1)^{(1-\kappa)\frac{n}{2}}\epsilon_{n+1}}{(n+1)^{\frac{d}{2}}}
\exp \left\{-n\pi
\frac{E_c}{E}\right\}\;,\\
&&\epsilon_n=\delta^{-1}\sqrt{\pi}\Psi\left(\frac{1}{2},-\frac{d-2}{2};
n\pi\frac{E_c}{E}\right).\nonumber
\end{eqnarray}
If $E/E_c<<1$ one can use an asymptotic of $\Psi$-function \cite{b12},
$\Psi\left(\frac{1}{2},-\frac{d-2}{2};
n\pi\frac{E_c}{E}\right)=\frac{1}{\sqrt{\pi
n}}\sqrt{\frac{E}{E_c}}+O\left(\left[\frac{E}{E_c}\right]^{-3/2}
\right).$ Then
$\delta=\sqrt{\frac{E}{E_c}},\;\epsilon_n=n^{-\frac{1}{2}}$ and $\mu=1$.
In this case the adiabatic field is equivalent to $T$-constant field
with the identification $\alpha=T\sqrt{\frac{E_c}{E}}$. At strong fields $E\sim
 E_c$  all the terms with different $\epsilon_n$ contribute to
the sum in (\ref{e20c}) and the expression for $P_v$  differs
essentially from the one for the $T$-constant field.

One can also remark that the case of a periodic alternating electric
field in 3+1 dimensions was also considered in the literature, for example,  
quasiclassically \cite{d4}, and  exactly \cite{d5}.

\subsection{ Constant field}

Here we consider the case of a  constant uniform electric field
(\ref{e4a}). Then   $E(x^0)=E$ and  potential $A_D=Ex^0$. 
Solutions of the eq.(\ref{e5}) in such a field can be found in the form  
\begin{equation}\label{e16}
{} _{+}^{-}\phi_{{\bf
 p},s}(x^0)=CD_{\nu -\frac{1+s}{2}}(\pm (1-i)\xi),\;\;{}_{-}^{+}\phi_{{\bf
 p},s}(x^0)=CD_{-\nu-\frac{1-s}{2}}(\pm (1+i)\xi). 
\end{equation}
Using an asymptotic expansion of  WPC-functions (\ref{e40ab}), 
one can get the  asymptotics of the quasi-energies,  
\[
{}_{\zeta}{\cal E}_{\bf
p}=\zeta |eEx^0-p_D|,\;\;{}^{\zeta}{\cal E}_{\bf p}=\zeta (eEx^0-p_D)\;,
\]
so that  IN and OUT-solutions can be constructed from (\ref{e16}) by
means of eq. (\ref{e6}). The same 
asymptotic expansion (\ref{e40ab}) allows one to calculate 
the normalization constants, $C=(2\pi )^{-D/2}
(2eE)^{-1/2}\exp\left\{-\pi\lambda/8\right\}$ for spinor case and $C=(2\pi )^{-D/2}
(2eE)^{-1/4}\exp\left\{-\pi\lambda/8\right\}$ for scalar one.
Straightforward calculations, similar to ones where made in the two previous
cases, lead to the  expression (\ref{e40ad}) for the mean numbers of
particles created. It does not depend on the dimensionality of the space
and coincides with the result which was derived in  \cite{b9a} for
3+1 QED.  In that paper the  authors 
used quasiclassical considerations to advocate the classification of
the solutions (\ref{e16}). The constant character of the field does not
allow one to treat consistently time divergences, so that they got
over them ``by hand'', using also quasiclassical considerations.  Now
one can see that the consideration of the $T$-constant field gives a
possibility both to ground all the results obtained from the constant field
solutions, solving consistently the problem of the  time divergences and
to go beyond the scope of the constant field to analyze the time scenario
of the process.

\subsection{Inclusion of a magnetic field}

In the same manner as before  one can  consider a more general case when a
constant uniform magnetic field is included, provided the invariant I is
negative. In fact, in $d>3$ there are $[\frac{d}{2}]-1$
 independent invariant parameters $H_j,\;j=1,2,\ldots
,H_{[\frac{d}{2}]-1}$ of the magnetic field, that corresponds to the 
possibility to construct $[\frac{d}{2}]$ invariants of the
electromagnetic field. In a convenient reference frame, the  magnetic part of the
field  tensor $F_{\mu\nu}$ is presented by the components 
$F^{\perp}_{\mu\nu}=\sum_{j=1}^{[\frac{d}{2}]-1}H_j(\delta^{j+1}_{\mu}\delta^{j}_{\nu}-
\delta^{j+1}_{\nu}\delta^{j}_{\mu}).$ One can always select solutions of
the squared Dirac equation (\ref{e3}) as eigenfunctions
for all independent nonzero  terms, which describe the interaction of intrinsic
magnetic moment of a particle with the external magnetic field. In
this case the matrices 
 $G\left({}_{\zeta}|{}^{\zeta '}\right)$ are diagonal and  one can
construct them using the
corresponding expressions in the pure electric 
field. Namely, for $d>3$ one has to make there a replacement
\begin{eqnarray}\label{e15} 
&&|{\bf p}_{\perp}|^2 \rightarrow  \sum_{j=1}^{[\frac{d}{2}]-1}\omega
_j+\omega_0,\;\;\;\omega_0
 =\left\{ \begin{array}{ll}
0, &\mbox{ d is even} \\
p^2_{d-2}, &\mbox{d is odd}\;,
\end{array}
\right. \\
&&\omega_j =\left\{ \begin{array}{ll}
 |eH_j|(2n_j+1-r_j),\;n_j=0,1,\ldots\;\;, & H_j\neq 0 \\
 p^2_{j}+p^2_{j+1}\;\;,&H_j=0
\end{array}\right.\;.\nonumber
\end{eqnarray}
In the presence of the magnetic field some momenta $p_j$ have to be
replaced by the discrete quantum numbers $n_j$. The number of these
momenta $p_j$ corresponds to the number of nonzero parameters
$H_j$. The magnetic field lifts the degeneracy in spin projections
in all the characteristics of the  particles creation effect.

We present here explicit formulas  in presence of the magnetic field  for  
the  total characteristics $N$ and $P_v$ in  case of the $T$-constant
field at big $T$,
\begin{eqnarray}\label{e19cc}
&&N= J_{(d)}^{\frac{1+\kappa}{2}}\frac{V_{(d-1)}Tm^2
\beta(1)}{2^{(d-1)}\pi^{d/2}}
\frac{E}{E_c}\exp\left\{-\pi\frac{E_c}{E}\right\}, \nonumber
\\
&&P_v=\exp\left\{-\mu 
N\right\},\;\; 
\mu=\sum_{n=0}^{\infty}\frac{(-1)^{(1-\kappa)\frac{n}{2}}\beta (n+1)}
{(n+1)\beta (1)}
\exp \left\{-n\pi
\frac{E_c}{E}\right\}\;,
\end{eqnarray}
where 
\begin{eqnarray*}
&&\beta(n)=
\prod_{j=1}^{(d-2)/2}\left\{\frac{eH_{j}}{\sinh (n\pi H_{j}/E)}
\left[\cosh(n\pi H_{j}/E)\right]^{\frac{1+\kappa}{2}}\right\}\;, \;\;d\;{\rm is\; even}\;,\\ 
&&\beta(n)=\left(\frac{m^2E}{n\pi E_c}\right)^{\frac{1}{2}}\;
\prod_{j=1}^{(d-3)/2}\left\{\frac{eH_{j}}{\sinh (n\pi H_{j}/E)}
\left[\cosh(n\pi H_{j}/E)\right]^{\frac{1+\kappa}{2}}\right\}\;,\;\;d \;{\rm is\;
odd}\;.
\end{eqnarray*}
The corresponding formulas for $3+1$-dimensional case were first
written in \cite{b9a,b9c}, and, in fact, can be derived easily from the
calculations of Schwinger \cite{b2}. They follow from (\ref{e19cc}) at
$d=4$. 

There exists a possibility to get also  exact results
for the particles creation in case
when a plane wave is added to the combination of electric and  magnetic fields. The
corresponding calculations for (3+1)-dimensional case were made in
cite{d6} for the electric field plus a plane wave field, and in
\cite{d7} for a general combination of electric, magnetic, and plane wave
fields. They can be generalized to any dimensions, combining the
approaches of \cite{d7} and of the present paper.

\section{Discussion}

\subsection{Time and space-dimensional analysis}

The calculation and analysis presented in Sect. III for fields, which are effectively
acting for a
finite time, allows one to study both pairs formation in time and  the
role of switching on and off effects. Besides, due to the fact that these
calculations are made in arbitrary dimensions of the Minkowski
space-time, one gets a possibility to  analyze the influence of the
dimensionality on the vacuum instability. 

Studying the T-constant field, one can see that the stabilization of
the mean numbers of particles created with given ${\bf p},r$
 in the form (\ref{e40ad}) for the longitudinal momenta
$|p_D|<<eE\frac{T}{2}$ comes at $T>>T_0$, where
$T_0=\frac{1}{\sqrt{eE}}(1+\lambda)
$. The characteristic time $T_0$ can
be called stabilization time. At the same time $N_{{\bf
p},r}$ for the big 
longitudinal momenta $|p_D|>>eE\frac{T}{2}$  decrease according to the power low
(\ref{e40abc}). 

The stabilization of the mean numbers  with given ${\bf p},r$
  in the adiabatic field in the
same form (\ref{e40ad}) comes for the longitudinal momenta
$|p_D|<<eE\alpha$
 at $\alpha >>\alpha_0$, $\alpha_0=\frac{1}{\sqrt{eE}}(1+\sqrt{\lambda})
$. For
big $|p_D|>>eE\alpha$  the mean numbers are exponentially small
(\ref{e19ba}).
For big $\alpha$ the adiabatic field varies slowly and coincides
nearly with the constant 
  one in the time interval $|x^0|\leq \alpha$. Then
$\alpha_0$ is a characteristic time of the stabilization in this field. 
 Thus, the stabilization
time $\alpha_0$ in the adiabatic field  differs from
the corresponding time $T_0$ in the T-constant field.
Thus, one can believe that the stabilization process depends of  the  switching on
and off  effects. In the case $E/E_c<1$, which corresponds to $\lambda >1,$  one can see
the stabilization comes quicker for adiabatic field than for the
T-constant one  ($\alpha_0<T_0$), i.e. the adiabatic  form of 
 switching on and off affects less the quantum system
than the instantaneous one in the T-constant field.  
If $E/E_c\geq 1$ , there exists a domain of
the transversal momenta ${\bf p}_{\perp}$ where $\lambda\leq 1$. In
 this case the stabilization times in both cases are the same, $\alpha_0\sim T_0\sim
\frac{1}{\sqrt{eE}}$, so that  for any $E$ the relation 
$\alpha_0\leq T_0$ holds.

Thus, one can conclude, that in some cases T-constant and adiabatic
electric fields act on the vacuum in a  similar way.
 However,  
the momentum dependence of the mean
numbers $N_{{\bf p},r}$  differs
essentially at big momenta for both fields. That is related to the 
switching on and off effects.
 To  estimate the role of the effects of
switching on and off  on the whole it is convenient
  to compare total
characteristics. First, let us compare the total mean numbers  with all
the longitudinal momenta, namely,  compare the formulas (\ref{e41ab}) and
(\ref{e20a}). In this case the effective action of both kind of fields is
the same if to identify $\alpha$ with $\sqrt{\lambda}T$. In spite of
this identification of $\alpha$ and $T$ is different for different
$\lambda$ (for different  ${\bf p}_{\perp}$), one can use it in a domain    
 ${\bf p}_{\perp}$ of the 
 transversal momentum,  $|\Delta {\bf p}_{\perp}|<< \sqrt{m^2+{\bf p}^2_{\perp}}$.
   As to the total numbers 
(\ref{e19}) and (\ref{e19c}), they
coincide  if one accepts the identification  $\alpha=T\delta^{-1}$. 
However, this identification
provides only the coincidence of $P_v$ for both cases (\ref{e20}) and
(\ref{e20c})  if $E/E_c<<1$ (then
$\delta=\sqrt{\frac{E}{E_c}})$. In this case the coefficients $ \mu$
in (\ref{e20}) and (\ref{e20c}) are the same.
 One can
conclude that the effects of switching on and off are not essential
for $E/E_c<<1$ and for big $T>>\frac{1}{m}\left(\frac{E_c}{E}\right)^{3/2}$, or
for big $\alpha$ respectively. In case of strong fields, 
$E/E_c\geq 1$,  these effects appear to be essential and  one has to take into
account the back reaction of particles created for more realistic
external field definition (e.g. see \cite{b5,b4,d1}).

The stabilization of the mean numbers of particles created with given
${\bf p},r$ at $T>>T_0$  can be  interpret in the following way: In
the T-constant electric field at the finite time instant $\frac{T}{2}$ the
pairs are created with quantum
numbers $|p_D|<eE\frac{T}{2}$ (equal for particles and
antiparticles). This corresponds to the region
$0<\pi^D(\frac{T}{2})<eET$ of the
observed kinetic momenta,
$\pi^D(\frac{T}{2})=-(p_D-eA_D(\frac{T}{2}))=-p_D+eE\frac{T}{2}$ 
of a particle  in each pair (direction of antiparticle kinetic momenta
is opposite). At $T>>T_0$ the effects of switching on
and off are already not essential. That is why the probabilities of
pairs creation do not depend on the time instants $t$,
 $-\frac{T}{2}<t<\frac{T}{2}$. One can think
that at this time instant the particles in the pairs are materialized
with almost zero 
longitudinal kinetic momenta at any given ${\bf p}_{\perp}$, i.e. with
the energies $\sqrt{m^2+{\bf p}^2_{\perp}}$. Then the
electric field accelerates them until the end of its action. Let us
suppose that   a
particle was created at a
time instant $t$. An expression for the kinetic momentum of such a
particle at the final time instant (which is equal to its expression
in the time instant when  the field switches off) can be found solving
the classical equation of motion $\frac{d\pi^D}{dx^0}=eE$, so that
$\pi^D(\frac{T}{2})=eE\left(\frac{T}{2}-t\right)$. Thus, a particle,
which was discovered with the quantum number $p_D$ at the final time
instant, was created at the time
instant $t=\frac{p_D}{eE}$. Then the integration over the longitudinal
momenta $p_D$ is equivalent to one over the time $t$,
$\int dp_D=eET$. This conclusion coincides with one derived in course
of the strict quantum analysis  presented in
Subsect. IIIA. According to the same interpretation, for particles
with relatively nonzero mean numbers,  the maximum value of the
kinetic momentum  $\pi^D(\frac{T}{2})$ is $eET$ that  corresponds 
to the particles, which were created at the initial time   
$t=-\frac{T}{2}$, whereas its minimal value is $0$ and corresponds 
to the particles, which were created at the final time instant 
$t=\frac{T}{2})$. This conclusion coincides also with one
derived from quantum consideration in Subsect. IIIA.

In the conclusion of the time analysis, one can remark that the time
$T_0$, which was introduced  by us as the stabilization time, was
interpreted in some papers as the time of a pair creation
\cite{b9a,b9ab}. However, we have seen that  in the
adiabatic field the stabilization time $\alpha_0$ is different, thus $T_0$ is not
an universal characteristic  and depends  of the field form.
In this connection one can propose another characteristic 
time, which  a pair formation in a quasiconstant electric field.
Indeed, as we have mentioned above,  
all the results in the  T-constant and adiabatic fields are comparable if
$E/E_c<<1$. In this
case  
the adiabatic  form of the  field is disturbing the 
quantum system less than the T-constant field. Here $\alpha_0<<T_0$ and $\alpha_0\approx
T_0^f=\frac{\sqrt{\lambda}}{\sqrt{eE}}$.
 Since the adiabatic field is
quasiconstant for the time interval $T^f_0$ and it is big enough for
the stabilization, one can 
interpret  $T^f_0$ as the time of a pair formation.
One can extrapolate this interpretation of $T^f_0$ for any field strength $E$.
A quasiclassical consideration confirms this interpretation. 
Thus,  a virtual
 particle with initial zero energy gets from the electric field for the time $T^f_0$ the
 energy $\sqrt{m^2+{\bf p}^2_{\perp}}$ necessary for the
materialization. It is easily to see that the time
$T_0^f$ is always  either less than the stabilization times $T_0,\alpha_0$
or equal to them. 
Some other consideration related to the time $T_0^f$ one can meet in the next subsection.

Turning to the dimensional analysis, one can see that the increase of
degrees of freedom, due to the increase of 
the dimensionality of the space-time itself and due to the related
increase of the spinning space dimension $J_{(d)}$, affects essentially
the total numbers of particles created in the unit of the volume and the
probability for   vacuum to remain a vacuum. 
Thus, the increase of spinning degrees of freedom leads to an increase of
$N$ and $P_v$ at any ratio $E/E_c$.
In particular, in $d>3$
the numbers of  fermions created is greater than the one of 
bosons. An increase of spatial dimensions leads to a decrease of the
total numbers
 of particles created in the unit of the volume and the
probability for a  vacuum to remain a vacuum at $E/E_c<1$ and their
increase at $E/E_c>1$.

 The presence of  walls or of a nontrivial topology affects the
spectrum of particles created. If the length $L_i$ of the space in the
direction of an axis $x^i$ is restricted by the walls 
 the corresponding momentum $|p_i|=\frac{2\pi n}{L_i},
\;n=1,2,\ldots \;\; $ is quantized.
At $L_i \sim \frac{1}{m}$ the dependence of the mean
numbers on the   boundary conditions is essential.
 At $L_i << \frac{1}{\sqrt{eE}}$ the mean
numbers $N_{{\bf p},r}$ in the quasistationary fields are very small
for any strength $E$. In this
connection one can treat $L_0= \frac{1}{m}[1+(\frac{E_c}{E})^{1/2}]$    
as  a characteristic dimension of the system, for which the boundary
conditions are essential.
 At $E/E_c\geq 1$ it is the Compton wave length.
It is interesting to remark that  the stabilization
times $T_0,\alpha_0$ coincide with $L_0$ at $E/E_c= 1$. 

Imposing
periodic conditions in the direction of an axis $x^i$ (that corresponds, in
particular, to the torus topology), one gets for the momentum
$|p_i|=\frac{2\pi n}{L_i},
\;n=0,1,2,\ldots\;\;. $ Then at $L_i << \frac{1}{\sqrt{eE}}$
 only particles with $p_i=0$ can be created. If the electric
field has the same direction, then the total number $N$ and the
probability $P_v$ do not depend on time $T$, since this dependence 
arises in course of a summation over the longitudinal momenta. It is
interesting that the presence of the magnetic field acts as a
dimensional reduction. Indeed, in the strong magnetic field with some $H_j>>E$ the
lowest energy level of a boson can not be less than $|eH_j|$, whereas for
a fermion it can. That means that the strong magnetic field acts on
bosons as some walls and on fermions as the presence of the torus
topology. Thus, one can see that if some of   the magnetic
fields are strong enough, then the corresponding  spin projection become frozen and
total characteristics, like total mean numbers  decrease.
 These dimensional effects may be relevant to the  matter
creation at early universe.

\subsection{Relation between the vacuum instability in external
electromagnetic and gravitational fields}

It is interesting to compare particles creation in  external
electromagnetic fields and in external fields of different nature, for
example, in external gravitational fields. To this end one can use
results obtained in the quasiconstant electric fields and in the
static gravitational fields. The latter problem was considered first
by Hawking \cite{b21} who, in particular, calculated the mean numbers
of particles created  by  static gravitational field of a black
hole with mass $M$ in a specific thermal environment,
\begin{equation}\label{ec1}
N_{n}=\left[\exp\left\{2\pi\frac{\omega}{g_{(H)}}\right\}+\kappa\right]^{-1},
\end{equation}
where $\omega$ is the energy of a particle created, which supposes to be
dependent on a complete set of quantum numbers $n$,
 $g_{(H)}= \frac{GM}{r^2_g}$, where $r_g$ is the gravitational radius,
so that $g_{(H)}$ is free falling acceleration at this radius.
This spectrum was interpreted as a Planck distribution  with the temperature
 $\theta_{(H)}=\frac{g_{(H)}}{2\pi k_B}$ ($k_B$ is the
Boltzmann constant).
 As before
$\kappa=+1$ for fermions and $\kappa=-1$ for bosons.
It is also known \cite{b22} that an observer, which is moving with a
constant acceleration $g_{(R)}$ (with respect to its proper time), will register
in the Minkowski vacuum  some particles (Rindler particles). The
mean numbers of  Rindler bosons have the same Planck form (\ref{ec1})
(with $\kappa =-1$), where one has to replace $g_{(H)}$ by $g_{(R)}$,
so that the correspondent temperature is $\theta_{(R)}=\frac{g_{(R)}}{2\pi k_B}$.
 One can
find many other examples when the particles creation in external
gravitation fields (and due to a nontrivial topology) can be described by
means of an effective temperature \cite{b5,b4}( see also references
in the recent publications \cite{b25}). On the other hand the distributions
obtained in external electromagnetic fields have not the thermal form
at a first glance. Nevertheless, there were attempts to find close
relations between the distributions in both cases, moreover, to derive
 Hawking distribution from the one in external electromagnetic
field.

In  \cite{b23a}, the distribution (\ref{e40ad}) at
${\bf p}_{\perp}=0$ was interpreted as the Boltzmann one for particles
in the ground state with the energy $m$ and the effective temperature
$\theta_{(E)}=\frac{2eE}{\pi m}$. The same temperature follows from some
other consideration \cite{b23b}  for the same restricted case.
Unfortunately, such an interpretation does not allow one to include 
 other states with nonzero momenta in the consideration.  

In  \cite {b24}, the authors did not  introduce an effective
temperature directly in the electrodynamical case but tried to find a 
relation between both distributions, in particular, to extract the
Hawking temperature from the electrodynamical formulas. We are going
to repeat briefly here this consideration , using some new details, which came
from the results of the present paper. As was established, a particle
with given momenta is created in a time instant with  energy
$\omega=\sqrt{m^2+{\bf p}^2_{\perp}}$, which corresponds to zero
longitudinal kinetic momentum at this time instant. Thus, namely this
expression plays the role of the total energy of the particle at the
time moment of creation. Then we can compare  equations of motion
for a classical particle in a constant electric field 
$d{\bf \pi}/dx^0= e{\bf E}$ with ones in the static
gravitational field  $d{\bf \pi}/dx^0= \omega
{\bf g}$. In the latter, $\omega$ is the total energy of the test
particle and ${\bf g}$ is the three-dimensional gravitational field
strength vector. Although these equations are formally similar, there
is a fundamental difference between them: the electromagnetic coupling
constant $e$ of a charged particle is not affected by its motion,
while the coupling to the gravitational field is proportional to the
total energy of the test particle. The latter property is a direct
consequence of the equivalence principle.
Let us formally replace the electric field strength $E$ by a quantity
that characterizes the gravitational field strength g and exploit
the equivalence principle by considering, that the coupling of the
particle to the field is proportional to the energy of the former,
which also allows us to replace $e$ by $\omega$. The expression that
arises from (\ref{e40ad}) as a result of these replacements can be
interpreted as the mean numbers of particles created by the
corresponding gravitational field and have the form of the Boltzmann
distribution with characteristic temperature 
$\theta'=\frac{{\rm g}}{\pi k_B}$. If g is the gravitational field
strength at the horizon of the black hole $g_{(H)}$, then $\theta'$ is
only 2 time greater  than the Hawking temperature $\theta_{(H)}$. 
In spite of the fact that an explicit progress was  achieved in the way of comparing  both
distributions, some questions remain. For example,
why the
temperature derived by means of the equivalence principle 
from the electrodynamical distribution differs by the factor 2
 from the Hawking one? Is there a thermal
interpretation of the electrodynamical distribution (\ref{e40ad}) or
some universal form of particles creation spectrum which is valid in
both cases? 
Below we propose some  interpretation of the electrodynamical formulas
which pretends to answer these questions.
We go beyond the classical consideration, taking into account
properties of the physical vacuum in the time dependent external
field.

 First, one can remark that due to the time dependence of the
potential $A_D(x^0)$, which defines the quasiconstant electric field,
the level of the vacuum energy  changes with time.
Thus, one has to calculate carefully the difference between the energies of
the system in the initial (vacuum) and in the final (with particles) states.
Let us  do the calculations  in the
case of fermions and in zero order with respect to radiative
corrections in the T-constant electric field. In this case the
corresponding Hamiltonian has the form
\begin{equation}\label{ec3}
H(x^0)=\int\bar{\Psi}(x)H_{o.p.}\Psi (x)d{\bf x}\;,
\end{equation}
where $H_{o.p.}$ was defined in (\ref{e8}), and $\bar{\Psi}(x),\;\Psi
(x)$ are electron-positron field operators in the generalized Furry
picture \cite{b15,b16,b6}. Being written in terms of IN- and OUT- operators of creation
and annihilation at $x^0\rightarrow\mp\infty$ respectively, the 
 Hamiltonian $H(x^0)$ has the diagonal forms,
\begin{eqnarray}\label{ec4}
&&H(x^0)=\sum_{{\bf p},r}p_0(t_1)\left[a^{\dagger}_{{\bf p},r}(in)
a_{{\bf p},r}(in)+b^{\dagger}_{{\bf p},r}(in)
b_{{\bf p},r}(in)-1\right],\;\;\;\;\;\;\;x^0\rightarrow -\infty, \nonumber\\
&&H(x^0)=\sum_{{\bf p},r}p_0(t_2)\left[a^{\dagger}_{{\bf p},r}(out)
a_{{\bf p},r}(out)+b^{\dagger}_{{\bf p},r}(out)
b_{{\bf p},r}(out)-1\right],\;\;x^0\rightarrow +\infty ,
\end{eqnarray}
where, as before, $t_2=-t_1=\frac{T}{2}$, and $p_0(t_i)
=\sqrt{m^2+{\bf p}^2_{\perp}+(\pi_D(t_i))^2}$ is the energy
of a particle in the initial and final time instants $t_i$ (the
longitudinal momenta $\pi_D(t_i)=p_D-eA_D(t_i)$ in the T-constant
field have the form $\pi_D(\pm\frac{T}{2})=p_D\mp eE\frac{T}{2}$.
Let us consider the variation of the total energy of the system, which
goes over from the initial vacuum state $|0,in>$ to the final state
$|N,out>$ with
pairs created in all the possible levels  for the T-constant field,
$$|N,out>=\prod_{{\bf p}_{\perp},r,|p_D|<eET/2}
a^{\dagger}_{{\bf p},r}(out)b^{\dagger}_{{\bf p},r}(out)|0,out>.$$
Then  one can formally write the energy of the initial state as
$${\cal E}_1=-\sum_{{\bf p},r}p_0(t_1),$$
and of the final state as
$${\cal E}_2=\sum_{{\bf p}_{\perp},r}\left[\sum_{|p_D|<eET/2}
2p_0(t_2)-\sum_{p_D}p_0(t_2)\right].$$
Thus
\begin{equation}\label{ec5}
\Delta {\cal E} ={\cal E}_2-
{\cal E}_1=\sum_{{\bf p}_{\perp},r}\left(\sum_{|p_D|<eET/2}
[p_0(t_2)+p_0(t_1)]+\Delta {\cal E}_{vac}\right),
\end{equation}
where 
\begin{equation}\label{ec6}
\Delta {\cal E}_{vac}=\sum_{|p_D|>eET/2}
[p_0(t_1)-p_0(t_2)]
\end{equation}  
is  the shift of the vacuum energy related to the levels
with given ${\bf p}_{\perp},r$, in which no pairs
appear. 
 We are going to analyze this shift only. That is
why we do not discuss here  regularization problems of the total sum
 (\ref{ec5}) (that can be done, using, for example, the methods described in
\cite{b19}).
One can see that
$$\sum_{|p_D|<eET/2}
[p_0(t_1)-p_0(t_2)]=0,$$
since $p_0(t_1)-p_0(t_2)$ is an odd function of $p_D$.
 That allows one to extend the summation in
(\ref{ec6}) over all the longitudinal momenta. The vacuum before the
time instant $t_1$ was free and therefore symmetric with respect to
the longitudinal kinetic momentum $\pi_D(t_1)=\pi_D=p_D+eE
\frac{T}{2}$. Replacing the summation over $p_D$ by one over $\pi_D$,
one can therefore treat the corresponding improper integral in
sense of its principal value. Thus 
\begin{eqnarray}\label{ec7}
\Delta {\cal E}_{vac}&=&\frac{L}{2\pi}\lim_{M\rightarrow\infty}
\int_{-M}^{M}\left(\sqrt{m^2+{\bf p}^2_{\perp}+\pi_D^2}-
\sqrt{m^2+{\bf p}^2_{\perp}+(\pi_D-
eET)^2}\right)d\pi_D  \nonumber\\
&=&-\frac{L}{2\pi}(eET)^2=-\frac{L}{2\pi}\left[\pi_D(t_2)-\pi_D(t_1)\right]^2.
\end{eqnarray}  
Since the number of  states with given $p_D$, in which particles can be
created, is equal to $\frac{1}{2\pi}eELT$, see (\ref{e41ab}), then the
shift (\ref{ec7})  can be  rewritten in the form
\begin{equation}\label{ec8}
\Delta {\cal E}_{vac}=\sum_{|p_D|<eET/2}
\Delta\epsilon_{vac},\;\;\Delta\epsilon_{vac}= -eET=-|\pi_D(t_2)-\pi_D(t_1)|.
\end{equation}  
Thus,
\begin{equation}\label{ec9}
\Delta {\cal E} 
=\sum_{{\bf p}_{\perp},r}\sum_{|p_D|<eET/2}\Delta\epsilon,\;\; \Delta\epsilon=
p_0(t_2)+p_0(t_1)+\Delta \epsilon_{vac}\;,
\end{equation}
where $\Delta \epsilon$ can be interpreted as a work which the external
field  accomplishes for the creation of a pair in a given state. It
contains a contribution $\Delta \epsilon_{vac}$ which takes into
account a shift of the vacuum energy in those states which remain
vacuum ones. 
The corresponding work with respect to a particle will be
denoted by $\omega$, so that, 
\begin{eqnarray}\label{ec9a}
\omega&=&\frac{1}{2}\Delta \epsilon \nonumber\\
&=&
\frac{1}{2}\left[\sqrt{m^2+{\bf p}^2_{\perp}+(\pi_D(t_2))^2}+
\sqrt{m^2+{\bf p}^2_{\perp}+(\pi_D(t_1))^2}-|\pi_D(t_2)-\pi_D(t_1)|\right]\;.
\end{eqnarray}
 Now we
remark that due to the conditions of the stabilization $T>>T_0$, $|p_D|<eE\frac{T}{2}$, under
which all the results for the T-constant field were obtained, we can write
\begin{equation}\label{ec10}
\omega=\frac{1}{4}\lambda eE
\left(\frac{1}{|\pi_D(t_2)|}+\frac{1}{|\pi_D(t_1)|}\right)
=
\frac{\lambda}{T}=\frac{\lambda eE}{2p_0(t_2)},\;\;
\lambda=\frac{m^2+{\bf p}^2_{\bot}}{eE}\;.
\end{equation}
Then the spectrum (\ref{e40ad}) can be rewritten in the following form
\footnote{we have restored $\hbar$ and $c$ here for
convenience of the reader}
\begin{equation}\label{ec11}
N_{{\bf p},r}=\exp\left\{-2\pi\frac{\omega}{\frac{\hbar}{c}g}\right\},
\end{equation}
where the quantity $g$ can be written in several equivalent forms  
\begin{equation}\label{ec11a}
g=\frac{c eE}{2}\left(\frac{1}{|\pi_D(t_2)|}+\frac{1}{|\pi_D(t_1)|}\right)
=\frac{2c }{T}=\frac{c eE}{p_0(t_2)}\;.
\end{equation}
The last expression in (\ref{ec11a}) allows one to treat $g$ as the
classical acceleration of a particle in the electric field  in the
final
 time instant $t_2=\frac{T}{2}$, for the case when  the
action time of the  field is big enough, so that the corresponding
velocities are near $c$. Formally this is valid under the quantum
condition of stabilization. The distribution (\ref{ec11}) is, in fact,
the Boltzmann one with the temperature 
$\theta=\frac{\hbar g}{2\pi c k_B}$ 
 having literally the Hawking form.
Thus, if one identifies the work $\omega$, we have introduced, with
the energy of a particle in the formula (\ref{ec1}),
then the distributions in electrodynamical and gravitational cases
have the same
thermal structure. Let us discuss now the possible origin of the
differences in the electrodynamical and gravitational formulas. First
of all, the formula (\ref{ec1}) 
 is derived in the formalism of the stationary scattering theory,
where it is not necessary to take separately into account a shift of the vacuum
level. In this case the energy of a particle created may
coincide with the corresponding work of the field. Second, the different
form of the thermal distributions (Boltzmann, Planck) can be
stipulated by essentially different situations in both cases. In the electrodynamical
case we deal in fact with pure states, whereas in the gravitational
problems a density matrix is arisen necessary due to the horizon of
events formation. At $\omega/g<<1$ the Planck spectrum coincides with
the Boltzmann one. In this case one can believe the form (\ref{ec11})
for the spectrum of particles created is universal and applicable to
any theory with quasiconstant external fields.

The form (\ref{ec11}) can be useful to describe situations in constant
fields, where it is convenient to avoid the consideration of the time
evolution, as we have seen comparing it with the gravitational
cases. Another form of the distribution (\ref{ec11}) with 
$2\omega=\Delta \epsilon$ from 
(\ref{ec9a}) (or (\ref{ec9})) and with the acceleration $g$ from
(\ref{ec11a}), 
can be useful in problems with explicit time dependence.

 The universality of the formula (\ref{ec11}) can be examine also in
the case of the adiabatic electric field, $\alpha>>\alpha_0$,
considered in Sect.IIIB. To apply it to the latter case one needs to
put $t_{2,1}\rightarrow\pm\infty$, then $\pi_D(t_2)=p_D-eE\alpha,\;
\pi_D(t_1)=p_D+eE\alpha$ and $p_0(t_2)=\omega_+,\;p_0(t_1)=\omega_-.$
 In this case according to (\ref{ec8})
$\Delta \epsilon_{vac}=-2eE\alpha$ and the distribution (\ref{e19a}) follows.

One can remark in this connection that in case $E/E_c<<1$
($\alpha_0<<T_0$) the formulas (\ref{ec10}) (and, therefore, (\ref{ec9a}))
in the adiabatic field are valid at the condition
$t_2=-t_1=\frac{T}{2}>>\alpha_0\sim T_0^f,$ which are weaker than in
the case of the T-constant field. That allows one to interpret the time
$T_0^f$, which already had appeared before in Subsect.A  as a pair formation
time, from another point of view. Considering the formula
(\ref{ec9a}), one can see that at $T>>T_0^f$ the $\Delta\epsilon$
is less than $2\sqrt{m^2+{\bf p}^2}$, due to the essential
contribution of the vacuum shift $\Delta \epsilon_{vac}.$ That means
the work of the external field to produce a pair is less than one,
which could be  
expected from the perturbation theory, where no vacuum change is taken
into account.

The consideration presented was made only for fermions. However, if
one believes that the quantity $\Delta \epsilon_{vac}$ can be taken
in the form (\ref{ec8}) for bosons as well, then the distribution
(\ref{ec11}) holds also in the scalar case. A consistent analysis for
charged boson is more complicated and needs to take into account
possible condensate formation and its evolution in an external field
(see, for example, \cite{b5,b28}).

Finally, we believe that the formulas derived and the
time-dimensional analysis presented can be also useful to
describe  some collective effects in the framework of quantum field
theory, for instance, to describe multiple particles creation by means
of the external field approach \cite{b26,d1} or in string models with external field
\cite{b19,b20,b27}.

\section{Acknowledgments}

D. Gitman  and S. Gavrilov thank Brazilian foundations   CNPq  
and  FAPESP respectively for support. Besides, S. Gavrilov thanks Russian 
Foundation of Fundamental Research which is
supporting him in part under the Grant No 94-02-03234. Both thank Prof. J. Frenkel for 
useful discussions and friendly support.

\end{document}